# RETHINKING VALUE CREATION FROM THE RESOURCE BASED VIEW: THE CASE OF HUMAN CAPITAL IN MOROCCAN HOTELS

YOUSSEF IFLEH, MOHAMED LOTFI and MOUNIME ELKABBOURI

ENCG /Hassan 1st University, Settat, Morocco

**Abstract**

The growth of the modern knowledge-based economy is becoming less and less dependent on tangible assets and more on intangible ones. In this context, the role of human capital in the value creation process has become central. Despite the large amount of scientific work on human capital phenomena, little research has revealed the role of human capital in the process of creating value. The purpose of this article is to evaluate the impact of human capital on value creation within 31 classified hotels in Morocco for the period 2013-2015. This paper is organized into four sections. First, we return to the main conceptualization of value creation. The goal of this first section is to synthesize prior work on this construct and highlight the main role of the resource based view (RBV) in explaining it. This view presents the point that links value creation to human capital given that this latter concept is one of the most resources of the firm. Next, we present the main definition of human capital. To do so, we make use of concepts from psychology, economy and strategic human resource management. Then, we shed light on the existing relationship between the two concepts of our research. Finally, we present the methodology of this research as well as the results. The required data to calculate value creation is obtained mainly from the annual reports of Moroccan hotels. Whereas, human capital is assessed by a questionnaire using the scale of Subramaniam and Youndt (2005). Data is examined using linear regression by PASW statistics software. The results of this study give a more concrete picture on the creation of value in this context and refute any link between these two concepts.

## I.        Introduction

The significant revolution in information and communication technologies, the emergence of knowledge economy and the importance of innovation as a determinant of competitiveness are changing the requirements of the market and leading organizations to face new challenges and risks. Indeed, the current environment has become increasingly complex, uncertain and highly aggressive. In this context, the role of human capital in the value creation process has become central. Quinn (1994) states that the modern firm derives its economic power from its intellectual capacities rather than its physical assets. The purpose of this article is to evaluate the impact of human capital on value creation in classified hotels in Morocco. To do so, we return, first, to the main definitions of both, human capital and value creation. Then, we shed light on the theoretical and empirical relationship between these two concepts. Finally we expose the methodology of this study as well as the results.

## II.    HUMAN CAPITAL: OMNIPRESENCE AND LACK OF CONSENSUS ON HIS DEFINITION

### 2.1.  Human capital: the birth of an economic concept

The first considerations on the importance of human capital go back to the 18th century, notably with the work of Adam Smith. The latter, in analyzing the causes of the wealth of nations, established the foundations of human capital conceptualization. For Smith, improving the skills of employees is a major source of economic development. What is essential to an economy are "the useful talents acquired by the inhabitants or members of society" and not the mass of the workers. He considers education and, more broadly, competences as a kind of capital which, for its constitution, requires a cost, but once acquired, it constitutes "a capital fixed and realized in its person".

Later, Alfred Marshall (1890) stressed the importance of investment in human beings. He pointed out the importance of education as a means of forming citizen's personality. It was not until the 1960s, with the work of Becker (1964) and Schultz (1963), to integrate this concept as such and to announce the birth of the human capital theory. These contributions made it possible to fill the gaps in theories of economic growth by highlighting the central role of HC.

### 2.2.  Human capital: The most important component of company's intellectual capital

Recent years have witnessed an increase in the importance of the role of intellectual capital for firms (Miles et al., 1998, Augier and Teece 2005). The *IC* concept has become very popular in scientific circles and has become the focus of many research and discussions. The first use of this term was attributed to John Kenneth in 1969. Despite the immense amount of research on the subject so far, there is no universally accepted and applied definition, despite some homogeneity in the studies (Kaufmann and Schneider 2004; Sullivan 2005). However, a three-dimensional classification has reached a certain degree of consensus: human capital, relational capital, and structural capital (Stewart 1998, Sullivan 1999, Kaufmann and Schneider 2004, Boedker et al 2005, Marr and Roos 2005). The proponents of this stream consider that the IC is intimately linked to the RBV as it's considered a potential source of competitive advantage for the company.

Despite the fact that HC is considered as a component among others of intellectual capital, it is recognized by several authors as the most important intangible resource (Johanson 2005, Marr and Roos 2005). Fornell (2000) contends that HC is the force that guides the other two components of the IC, namely, relational capital and social capital. Moreover, the development of a structural capital is conditioned by the existence of the HC. Besides, the studies which have dealt with the interaction between the elements of IC have proven the supremacy of HC. Bontis et al (2000) confirmed that HC positively affects the company's structural capital. Moon and Kym (2006) and Hsu and Fang (2008) have agreed on the positive role of HC in the development of other IC components

### 2.3. Human capital in differential psychology

HC Studies conducted in a psychological context were concerned in most cases with personality traits and cognitive abilities. Defenders of this stream consider that the origin of the HC resource depends on the individual level and are found in the KSAOs (Knowledge, Skills, abilities and other characteristics " (Ployhart, 2011). KSAO constitute a subset of psychological differences that are relatively stable over time. According to Gerhart and Wright (2006), "knowledge" is the factual or procedural information needed to perform a specific task and is the basis on which skills and abilities are developed; "Skills" refers to the individual level of abilities to perform a specific task; The "abilities" correspond more generally to the capacity necessary for an individual to carry out the tasks related to a function; Finally, "other characteristics" often refer to personality traits or other attributes that affect individual ability to perform a specific task (Ployhart, 2011). Accordingly, individuals are supposed to have heterogeneous abilities based on their KSAOs which are generally stable and having an intra-psychic origin. The researchers examined how these KSAOs influence their productivity (Schmidt & Hunter, 1998).

### 2.4. Human capital in strategic human resource management (SHRM)

SHRM aims for linking a firm's employees to strategic needs (Wright and McMahan, 1992), using multi-theoretical, multi-level, and multidisciplinary processes (Wright & McMahan, 2011). However, the main theory that leads the strategic perspective of HRM is the RBV (Resource Based View) (Barney, 1991). This field developed to provide a theoretical basis for HR research (McMahan et al., 1999), using several theories, to mention only the most important ones: The RVB (Barney, 1990), the theory of human capital (Becker, 1964), agency theory (Jensen & Meckling, 1976), behavioral approach (Jackson, Schuler and Rivero, (Williamson, 1975) and dynamic capacities (Teece, Pisano & Shuen, 1997). However, after years of research, there seems to be a consensus regarding the supremacy of RBV. According to McMahan, et al. (1999), the RBV becomes the most dominant theory in the field of SHRM. It is widely used both in the development of theoretical studies (Huselid 1995, Lado & Wilson 1994, McMahan et al., 1999) and in empirical research (Combs, Liu, Hall & Ketchen, 2006; Harris , McMahan & Wright, 2009). According to Nyberg et al (2014), the SHRM is the result of the convergence of HRM and RBV. This approach assumes that the use of internal resources generates a sustainable competitive advantage (Barney, 1991) and among these resources are human resources that are considered potentially strategic for the company (Becker & Gerhart, 1996). The work on human capital resource in SHRM considers the latter to be a source of developing competitive advantage if it respects four conditions: it must be valuable within the firm, rare, inimitable and non-substitutable in labor market and other firms (Barney, 1991; Wright & Barney, 1998). Hence, from this perspective, the HC is apprehended at an organizational level as a strategic resource without  taking into account that this resource is more widely developed and accumulated within the framework of a specific working group (team or department for example) (Ployhart & Moliterno, 2011, Ployhart et al., 2011).

## III.        FROM VALUE TO VALUE CREATION

### 3.1.   The concept and history of value

The value notion has been studied in several disciplines: human resources, marketing, entrepreneurship, strategic management, psychology, sociology,...), and has been approached by several theories: Value theory (Porter 1985), resource-based theory (Barney 1991), transaction costs theory (Williamson 1975), stakeholder theory (Freeman 1984). This subject arouses enthusiasm on the part of several researchers and authors. This is reflected, on the ground, by the proliferation of several publications relating to this theme. This polysemy explains to a large extent its richness and the difficulty of associating it with a precise definition.

In economics, this concept has long been the subject of several debates in the mainstream of thought. Classics have developed objective theories of value. The aim is to develop objective and valid criteria for determining the value of everything at any time and in any place. Thus, Adam Smith distinguished between use value and exchange value. The use value corresponds to the utility of a good, whereas the exchange value concerns value of an asset with the amount of work it commands. Later, David Ricardo introduced a novelty in relation to Smith's theory by advancing the notion of "labor embodied" which, in addition to the quantity of labor necessary to produce an asset, included the quantity of labor embodied in the tools of production and intermediate consumption. Despite this difference, the two economists adopt the same logic, namely, the value of a good is determined by the amount of labor it contains.

It seems clear that despite the divergences in the assumptions, the defenders of the classical current agree that the value of a good is inseparable from work. These early attempts to explain value emphasize the importance of human involvement in this process and give us a signal on the dependence of the value on the human being.

As from the years 1870s, a new trend emerged, namely neo-classical theory. The latter indicates that the value of a good depends not only on its intrinsic qualities, but also on the circumstances. The proponents of this current are the marginalists in reference to the principle of marginal utility. Indeed, this principle specifies that the exchange value of an asset is dependent on its marginal utility. The latter depends, in turn, on the rarity and subjective tastes of individuals. Henceforth, several subjective theories have emerged, thus announcing the arrival of the neoclassical economy and a passage from theories that base value on utility to those based on labor.

The debate on the value theory in economics has always led to totally opposite views of society and represents a philosophical, ideological and political challenge. This discussion will not be prolonged as we consider it to be of no use to our research. We retain that the value of labor has a fundamental role in the economy.

### 3.2. Value creation: from traditional vision to cognitive vision.

The traditional financial vision comes from neoclassical theory. It implies that the creation of value is manifested by the existence of a surplus after having paid capital providers. The latter are the only ones entitled to this value thus created. This approach has been called into question with the advent of the contractual vision (Berle and Means, 1932, Jensen and Meckling, 1976, Fama, 1980, Fama and Jensen, 1983). This view highlighted the impact of managers in this process. Indeed, the separation between capital and management has led to interest conflicts and an information asymmetry between managers and shareholders. In such situation, capital providers were obliged to introduce internal and external control mechanisms to encourage management to act in their favor. This control resulted in additional costs, called agency costs.

A second reconsideration of the value creation theory was carried out by entrenchment theory (Shleifer and Vishny, 1989; Castanias and Helfat, 1992). Entrenchment is defined as a voluntary of manager to neutralize the control mechanisms which are imposed by the principal; what to allow granting itself more important personal advantages (Walsh and Seward 1990).

A third questioning was carried out by adopting the stakeholder theory which is a pluralistic (or partnership) vision oriented mainly by the contributions of Blair (1995), Charreaux and Desbrieres (1998). This vision challenges the simplistic role attributed to stakeholder (apart from capital providers) within the shareholder framework. The partnership approach gives to all stakeholders the responsibility of a residual risk and not only to the shareholders. Consequently, it will be justified to grant them a share of the created value.

Despite the differences between the stakeholder approach and the shareholder one, they both have one thing in common: the same original approach, in this case the disciplinary approach, a legal and financial vision whose objective is to ensure the distribution that maximizes value and minimizes conflict. The discipline that is applied to the managers tries to oblige them to make the best possible choices.

Works that adopt a cognitive and behavioral vision announce the beginning of a fourth stage. This vision states that cognitive levers such as learning skills and innovation are the basis for creating value. This vision is based on skills and resources. Thus, the way in which the value created is distributed is no longer a concern; however, it is the way in which this value is generated which represents the central stake of this vision. Cognitive theories are mainly based on four currents. The first is the behavioral approach (Simon, 1947, March and Simon, 1958, Simon and March, 1963). This takes into account the behavioral bias which constitutes the gap between the actual behavior of the individuals and the behavior supposed to maximize the expectation of utility. Its objective is to explain non-rational decisions taken by managers. This approach has enriched the visions mentioned earlier by examining the effect of these biases on disciplinary and cognitive levers as well as on the conflicting relationships between the different stakeholders.  The second approach relates to organizational learning (Argyris and Schon, 1978) which addresses the role of learning cognitive organizations in order to enhance the competitiveness and performance of the organization. The third stream is based on resource-based theories (Penrose, 1959, Prahalad and Hamel, 1990, Barney, 1991). This theory considers that the competitive advantage of an organization comes essentially from its own resources. The fourth stream is based on neo-Schumpeterian evolutionary economic theory (Nelson and Winter, 1982). The latter considers the firm as an entity that encompasses productive knowledge and an interpretative system favoring the notion of competition based on innovation. This research is part of the resource-based theory that justifies the central role of the human capital resource in the process of value creation.

## IV.        HUMAN CAPITAL AND VALUE CREATION

The literature underlines that the greatest challenge for human capital researchers is to prove that the latter creates value (Česynienė & Stankevičienė; 2011). There is no universal method for assessing the value created by human capital and showing it in the form of numbers or graphs. In the 1960s and 1970s, financial methods and indicators dominated the assessment of the contribution of human capital to value creation. However, to the best of our knowledge, no studies have attempted to link human capital directly to the creation of financial value. On the other hand, several attempts that evolve in this direction have taken place. Similarly, Beattie, Smith (2010) argues that some authors report the effectiveness of investments in human capital to the competitive advantage of the organization. Guthrie and Petty (2007) in their empirical study, highlighted six components of human capital: know-how, education, occupational skills, work-related knowledge, professional skills and entrepreneurship. The aim was to elucidate whether these components were perceived as an antecedent of value creation and whether they were reflected in the annual reports of Australian firms. Brennan and Bozzolan (2008) analyzed the same components in 11 Irish companies and 30 Italian companies. Beattie and Smith (2010) undertook an empirical study of the contribution of human capital to value creation. The authors carried out a comparative analysis of the role of human capital in the generation of the value of the company according to 160 functional managers: 67 human resources directors and 93 financial directors.

Based on these data, the central hypothesis of the study is as follows: Human capital positively influences value creation in Moroccan hotels.

## V. METHODOLOGY AND SAMPLE CHOICE

### 5.1. The study sample

Our sample is selected according to the quota method. This is a non-probabilistic method which allows obtaining a sample with a certain representativeness of the studied population (Thiétart, 2007). It consists of segmenting the mother population according to predefined criteria. In the present study we will adopt two criteria: the distribution of overnight stays by destination (Table1) and the distribution of the classified hotels according to their category (Table 2).

The sample size is conditioned, among others, by the statistical method used. In this sense, Thiétart et al (2007) confirm that its determination is facilitated, in the case of using averages and frequencies in statistical analysis, by the existence of simple formulas. On the other hand, these formulas are not available in the case where the methods used are more complex, such as regression. These authors suggest inspiring from similar studies to imitate them. To do this, we have identified studies that have adopted similar approaches. Their analysis shows that the authors used samples of different sizes. This removes any obstacles to the choice of size. However, given the means available to us and the fact that statisticians consider that more than 30 is a large sample, we decided to conduct our study on 31 hotels.

The sample distribution is made, first, according to the two criteria already chosen. Table 3 details our sample. The final sample does not include hotels located in the cities of Ouarzazate, Essaouira, Tetouan and others. This choice is justified by the low concentration of hotels in these cities. It's worth mentioning that questionnaires are filled out by hotel managers or human resources directors. Financial information are extracted from the financial statements of years 2013, 2014 and 2015. This study proceeded over one five months-period.

| Destination | % Night |
|---|---|
| Marrakech | 34% |
| Agadir | 26% |
| Casablanca | 10% |
| Tanger | 5% |
| Fès | 4% |
| Rabat | 3% |
| Ouarzazate | 2% |
| Essaouira | 2% |
| Tétouan | 2% |
| Other | 12% |

**TABLE 1: DISTRIBUTION OF CLASSIFIED HOTELS IN MOROCCO ACCORDING TO DESTINATION (MINISTRY OF TOURISM, 2014)**

| Class | 1 Star | 2 Stars | 3 Stars | 4 Stars | 5 Stars | Total |
|---|---|---|---|---|---|---|
| Number | 270 | 212 | 213 | 187 | 77 | 959 |
| % | 28,15% | 22,11% | 22,21% | 19,50% | 8,03% | 1 |

**TABLE 2: DISTRIBUTION OF CLASSIFIED HOTELS IN MOROCCO ACCORDING TO THEIR CATEGORY (MINISTRY OF TOURISM, 2014)**

| | | 2 Stars | 3 Stars | 4 Stars | Total |
|---|---|---|---|---|---|
| | | 34,65% | 34,8% | 30,55% | 100% |
| Marrakech | 43% | 4 | 4 | 4 | 12 |
| Agadir | 33% | 4 | 4 | 3 | 11 |
| Casablanca | 13% | 1 | 1 | 1 | 3 |
| Tanger | 6% | 1 | 1 | 1 | 3 |
| Fès | 5% | 1 | 1 | 0 | 2 |
| Total | 100% | 11 | 11 | 9 | 31 |

**TABLE 3: DISTRIBUTION OF THE STUDY SAMPLE**

### 5.2. Operationalization of concepts

**Human capital**

This study uses Subramaniam and Youndt (2005) scale to evaluate human capital. This scale, which is also based on the perspective of the field manager, is inspired by the work of Lepak and Snell (2002). The collective HC is measured using a 5-point Likert scale. (1 = strongly disagree, 5 = strongly agree). This tool is used in several recent works (Aryee et al., 2013, Cabello-Medina et al., 2011, Liao et al., 2009, Takeuchi et al.

**Value creation**

In this study we use the Economic Value Added (EVA) to assess the value creation of hotels. It is one of the most popular methods in the financial press that was designed by the consulting firm Stern Stewart. The EVA is the result of the difference between NOPAT (Net operating profit after tax) and the return on invested capital (Equity and debt)

The $\beta i$ (Beta) is generally obtained by comparing the return rate of a share and that of the market. The determination of this coefficient for the hotels chosen in our sample poses a problem related to the fact that they are not listed on the stock exchange. To deal with this problem, the French evaluation society proposes the calculation of an "Unlevered" Beta, which consists of calculating the ß of a comparable company and neutralizing it by considering it outside its financial structure.

## VI. RESULTS AND ANALYSIS

The results of this study are presented as follows:

**Human capital**

The human capital items vary between 2.18 and 2.89. That said, the respondents generally disagreed with the questions put to them. This reflects a low estimate of the human potential available in the hotels sample.

**Value creation**

The EVA values are very heterogeneous with a coefficient of variation of 240.456% and a positive mean which contends that hotels create value.

**Human capital and value creation**

This model aims to establish the link between the human capital assessed at the hotel level and the value creation measured by the EVA. The authors who used the scale of Youndt, M.A., Subramaniam, M., & Snell (2005) to evaluate human capital, have always calculated the average of the five components . Therefore, we will test the relationship between the mean given by the five components of the human capital scale and the value creation measured by the EVA. The results given by the software are as follows:

| Correlations | | | |
|---|---|---|---|
| | | EVA | Hotel HC |
| **Pearson Correlation** | EVA | 1,000 | -,450 |
| | Hotel HC | -,450 | 1,000 |
| **Sig. (unilateral)** | EVA | . | ,006 |
| | Hotel HC | ,006 | . |
| **N** | EVA | 31 | 31 |
| | Hotel HC | 31 | 31 |

**Table 4: Correlation coefficients of the model.**

| Model | | R | R-square | Adjusted R-square | Std error of the estimate | Changement dans les statistiques | | | | |
|---|---|---|---|---|---|---|---|---|---|---|
| | | | | | | R-square Variation | F Variation | df1 | df2 | Sig. Variation F |
| dimension0 | 1 | ,450a | ,203 | ,175 | ,0362188678 | ,203 | 7,373 | 1 | 29 | ,011 |
| a. Predictors : (constant), Hotel HC | | | | | | | | | | |
| b. dependant Variable : EVA | | | | | | | | | | |

**Table 5: Estimation coefficients of the model**.

Computer output shows that the dependent variable (EVA) is significantly and inversely correlated with the explanatory variable HC. Contrary to what was expected, HC of the hotels is not positively linked to value creation in this context.

The adjusted R-2 given by the software is 0, 175. This means that the variable (HC) accounts for 17.5% of the EVA variation.

To test whether the HC has a positive effect on the value creation (H1: p> 0, unilateral test on the right), the value of T (given by software) must be greater than the value given by the Student table at ($\alpha$/2 and n-2). In our case the value read in the table is 2.4620 and the value from the computer output is -2.715. This result leads us to accept H0, which means that the effect is not positive. Table 5 presents the estimation coefficients of the model.

## VII.        CONCLUSION

Value creation is a very important subject in management sciences. The cognitive approach was a change in how to design this concept by placing more emphasis on cognitive levers such as learning skills and innovation. As such, the resource-based view provides an appropriate framework for linking the resources available to the organization with value creation. The results of this study express a low estimate of the human potential available in Moroccan hotels. Moreover, value creation is characterized by a strong dispersion. In addition, contrary to expectations, human capital has no positive influence on value creation. Indeed, this result is not surprising given that several authors confirm that operational performance measures approach human capital more strongly than the overall measures of performance (Crook et al., 2011). From this perspective (Ray et al., 2004) suggest that if the research only considers overall performance measures such as EVA and fails to measure the results of operational performance, significant resources of competitive advantages Could be hidden. Studies using this approach suggest that operational measures should be taken into account. Crook et al (2011) propose the integration of these variables as mediator variables. In the case of hotels, the Revpar (Revenue by Available Room) is one of the most important measures of operational performance. For this reason, we recommend the use of this indicator in future research.

Moreover, these results challenge us on the financial situation of Moroccan hotels, as  many of them destroy value. Thus, we emphasize the importance of treating this relationship by considering a larger sample to verify its veracity.

Thus, we emphasize an important contribution of this study, which consists in using an internationally recognized scale of human capital, as well as the estimation of value creation using the EVA method using the unlevered Beta. To our knowledge, this is a first in the Moroccan context.